\documentclass[10pt]{article}
\begin{document}
\newcommand{\noi}{\noindent}
\newtheorem{opgave}{Excercise}
\newcommand{\bop}[1]{\begin{opgave}\begin{rm}{\bf#1} \newline\noi}
\newcommand{\eop}{\end{rm}\end{opgave}}

\newcommand{\nc}{\newcommand}
\nc{\dia}[1]{\mbox{{\Large{\bf diag}}}[#1]}
\nc{\diag}[3]{\raisebox{#3cm}{\epsfig{figure=diagrams/d#1.eps,width=#2cm}}}
\nc{\bq}{\begin{equation}}
\nc{\eq}{\end{equation}}
\nc{\bqa}{\begin{eqnarray}} 
\nc{\eqa}{\end{eqnarray}}
\nc{\nl}{\nonumber \\}
\nc{\grf}{Green's function}
\nc{\grfs}{Green's functions}
\nc{\cgrf}{connected Green's function}
\nc{\cgrfs}{connected Green's functions}
\nc{\f}{\varphi}
\nc{\exb}[1]{\exp\!\left(#1\right)}
\nc{\avg}[1]{\left\langle #1\right\rangle}
\nc{\suml}{\sum\limits}
\nc{\prodl}{\prod\limits}
\nc{\intl}{\int\limits}
\nc{\ddv}[1]{{\partial\over\partial #1}}
\nc{\ddvv}[2]{{\partial^{#1}\over\left(\partial #2\right)^{#1}}}
\nc{\la}{\lambda}
\nc{\eqn}[1]{Eq.(\ref{#1})}
\nc{\eqns}[1]{Eqs.(\ref{#1})}
\nc{\appndix}[3]{\section{Appendix #1: #2}\input{#3}}
\nc{\cala}{{\cal A}}
\nc{\calb}{{\cal B}}
\nc{\cale}{{\cal E}}
\nc{\calf}{{\cal F}}
\nc{\calh}{{\cal H}}
\nc{\calc}{{\cal C}}
\nc{\cald}{{\cal D}}
\nc{\calp}{{\cal P}}
\nc{\calfi}{{\cal\Phi}}
\nc{\lra}{\;\leftrightarrow\;}
\nc{\stapel}[1]{\begin{tabular}{c} #1\end{tabular}}
\nc{\al}{\alpha}
\nc{\be}{\beta}
\nc{\g}{\gamma}
\nc{\ka}{\kappa}
\nc{\om}{\omega}
\nc{\si}{\sigma}
\nc{\ro}{\rho}
\nc{\vph}{\vphantom{{A^A\over A_A}}}
\nc{\wph}{\vphantom{{A^A}}}
\nc{\order}[1]{{\cal O}\left(#1\right)}
\nc{\De}{\Delta}
\nc{\de}{\delta}
\nc{\vx}{\vec{x}}
\nc{\vxi}{\vec{\xi}}
\nc{\vy}{\vec{y}}
\nc{\vk}{\vec{k}}
\nc{\vq}{\vec{q}}
\nc{\vp}{\vec{p}}
\nc{\lijst}[1]{\begin{center}
  \fbox{\begin{minipage}{12cm}{#1}\end{minipage}}\end{center}}
\nc{\lijstx}[2]{\begin{center}
  \fbox{\begin{minipage}{#1cm}{#2}\end{minipage}}\end{center}}
\nc{\eps}{\epsilon}
\nc{\caput}[2]{\chapter{#1}\input{#2}}
\nc{\pd}{\partial}
\nc{\dtil}{\tilde{d}}
\nc{\vskp}{\vspace{\baselineskip}}
\nc{\coos}{co\-ord\-in\-at\-es}
\nc{\calm}{{\cal M}}
\nc{\calj}{{\cal J}}
\nc{\call}{{\cal L}}
\nc{\calt}{{\cal T}}
\nc{\calu}{{\cal U}}
\nc{\calw}{{\cal W}}
\nc{\msq}{\langle|\calm|^2\rangle}
\nc{\nsym}{F_{\mbox{{\tiny{symm}}}}}
\nc{\dm}[1]{\mbox{\bf dim}\!\left[#1\right]}
\nc{\fourv}[4]{\left(\begin{tabular}{c}
  $#1$\\$#2$\\$#3$\\$#4$\end{tabular}\right)}
\nc{\fs}[1]{/\!\!\!#1}
\nc{\dbar}[1]{{\overline{#1}}}
\nc{\tr}[1]{\mbox{Tr}\left(#1\right)}
\nc{\row}[4]{$#1$ & $#2$ & $#3$ & $#4$}
\nc{\matv}[4]{\left(\begin{tabular}{cccc}
   #1 \\ #2 \\ #3 \\ #4\end{tabular}\right)}
\nc{\twov}[2]{\left(\begin{tabular}{cccc}
   $#1$ \\ $#2$\end{tabular}\right)}
\nc{\ubar}{\dbar{u}}
\nc{\vbar}{\dbar{v}}
\nc{\lng}{longitudinal}
\nc{\pol}{polarization}
\nc{\longpol}{\lng\ \pol}
\nc{\bnum}{\begin{enumerate}}
\nc{\enum}{\end{enumerate}}
\nc{\nubar}{\overline{\nu}}
\nc{\mau}{{m_{\mbox{{\tiny U}}}}}
\nc{\mad}{{m_{\mbox{{\tiny D}}}}}
\nc{\qu}{{Q_{\mbox{{\tiny U}}}}}
\nc{\qd}{{Q_{\mbox{{\tiny D}}}}}
\nc{\vau}{{v_{\mbox{{\tiny U}}}}}
\nc{\aau}{{a_{\mbox{{\tiny U}}}}}
\nc{\aad}{{a_{\mbox{{\tiny D}}}}}
\nc{\vad}{{v_{\mbox{{\tiny D}}}}}
\nc{\mw}{{m_{\mbox{{\tiny W}}}}}
\nc{\mz}{{m_{\mbox{{\tiny Z}}}}}
\nc{\mh}{{m_{\mbox{{\tiny H}}}}}
\nc{\gw}{{g_{\mbox{{\tiny W}}}}}
\nc{\qw}{{Q_{\mbox{{\tiny W}}}}}
\nc{\qc}{{Q_{\mbox{{\tiny c}}}}}
\nc{\law}{{\Lambda_{\mbox{{\tiny W}}}}}
\nc{\obar}{\overline}
\nc{\guuh}{{g_{\mbox{{\tiny UUH}}}}}
\nc{\gwwz}{{g_{\mbox{{\tiny WWZ}}}}}
\nc{\gwwh}{{g_{\mbox{{\tiny WWH}}}}}
\nc{\gwwhh}{{g_{\mbox{{\tiny WWHH}}}}}
\nc{\gzzh}{{g_{\mbox{{\tiny ZZH}}}}}
\nc{\ghhh}{{g_{\mbox{{\tiny HHH}}}}}
\nc{\ghhhh}{{g_{\mbox{{\tiny HHHH}}}}}
\nc{\gzzhh}{{g_{\mbox{{\tiny ZZHH}}}}}
\nc{\thw}{\theta_W}
\nc{\sw}{{s_{\mbox{{\tiny W}}}}}
\nc{\cw}{{c_{\mbox{{\tiny W}}}}}
\nc{\ward}[2]{\left.#1\right\rfloor_{#2}}
\nc{\vpa}{(1+\g^5)}
\nc{\vma}{(1-\g^5)}
\nc{\gwzh}{{g_{\mbox{{\tiny WZH}}}}}
\nc{\gudh}{{g_{\mbox{{\tiny UDH}}}}}
\nc{\gwcj}{{g_{\mbox{{\tiny Wcj}}}}}
\nc{\gwcjg}{{g_{\mbox{{\tiny Wcj$\g$}}}}}
\nc{\gwwcc}{{g_{\mbox{{\tiny WWcc}}}}}
\nc{\gzcc}{{g_{\mbox{{\tiny Zcc}}}}}
\nc{\none}[1]{ }
\nc{\Vir}{{V_{\mbox{{\small IR}}}}}

\begin{center}
{\bf{\Large Majoranized Feynman rules}}\\  \vspace{\baselineskip}
R. Kleiss\footnote{{\tt R.Kleiss@science.ru.nl}},
I. Malamos\footnote{{\tt J.Malamos@science.ru.nl}}
 and G. v.d. Oord\footnote{{\tt vdoord@nikhef.nl}}\\
IMAPP, FNWI\\ Radboud University, Nijmegen, the Netherlands\\
\vspace*{5\baselineskip}
{\bf Abstract}\\ \vspace{\baselineskip}
\begin{minipage}{10cm}{
We point out that the compact Feynman rules for Majorana fermions proposed by
Denner {\it et al.\/} are in fact a convention for the complex phases
of (anti)spinors, valid for both Majorana and Dirac fermions. 
We establish the relation of this phase convention with
that common in the use of spinor techniques.}\end{minipage}
\end{center}
\vspace{3\baselineskip}

\setcounter{footnote}{0}
In \cite{denner}, compact Feynman rules for Majorana fermions have been proposed
that allow simple computations in for instance supersymmetric theories. These are
based on a distinction between the {\em fermion number flow\/} along a fermion line,
always taken to follow the direction of the arrow on the fermion line, and an
arbitrarily assigned {\em fermion flow\/} that can be taken in the direction of the
arrow or against it. Majorana fermions, lacking a definite fermion number, of course
have only a fermion flow. The assignment of (anti)spinors for external fermions in any
process, as well as the fermion propagators, are then based on the fermion flow.
For instance, an incoming electron with momentum $p$ is assigned the spinor
$u(p)$ if the fermion flow follows the fermion number flow (into the diagrams) ; 
alternatively, it is assigned $\vbar(p)$ if the fermion flow is chosen to be opposite
to the fermion number flow.  The Feynman rules for the various vertices
also pick up some minus signs depending on the relative orientation of
fermion flow and fermion number flow. It is obvious that the
possibility of formulating such Majoranized Feynman rules must be independent
of whether the theory actually contains Majorana fermions or only Dirac
fermions.

The method described in \cite{denner} is based on
the use of the charge conjugation matrix $C$ to relate $u$ and $v$ :
\bq
u = C\left(\vbar\right)^T\;\;,  \label{dennereqn}
\eq
where ${}^T$ denotes transposition. Now, two remarks are in order. In the first
place, the {\em only\/} properties that Dirac matrices have to obey are
\bq
\{\g^\mu,\g^\nu\} = 2\,g^{\mu\nu}\;\;\;,\;\;\;
\left(\g^0\right)^\dag = \g^0\;\;\;,\;\;\;\left(\g^k\right)^\dag = -\g^k\;\;(k=1,2,3)\;\;,
\eq
and any scattering amplitude should be (up to an overall phase) be independent
of the particular representation chosen for the Dirac matrices : in contrast,
the precise form of $C$ in terms of the $\g^\mu$
depends sensitively on the particular
representation. This lack of elegance leads us to state that {\em any\/} result
for scattering amplitudes obtained
by invoking the charge conjugation matrix $C$ must also be provable without any
referral to $C$ at all\footnote{Furry's theorem is an example : in
\cite{pppp} a proof without $C$ is given.}. In the second place, 
the Dirac spinors are defined by
their projection operators, which we take to 
be\footnote{There is some arbitrariness
in how the spin vector $s$ ought to transform if a spinor 
$u$ is transformed into
an antispinor $v$. This depends on whether one 
takes $s$ to refer to the particle's
angular momentum or to its magnetic momentum. We shall simply take $s$
to be defined from \eqn{projectors}.}
\bqa
&& u(p,s)\ubar(p,s) = \Pi(m,p,s)\;\;\;,\;\;\;v(p,s)\vbar(p,s) = \Pi(-m,p,s)\;\;\;,\nl
&&\Pi(m,p,s) = {1\over2}\left(\wph\fs{p}+m\right)\left(1+\g^5\fs{s}\right)\;\;.
\label{projectors}
\eqa
The distinction between particle and antiparticle is simply in the sign of $m$.
The only thing left undetermined by these definitions is the complex phase of
the spinors. If an incoming electron is always written as $u$, 
this overall complex phase
is unimportant ; but as soon as in some other diagrams the incoming 
electron is also written as $\vbar$, the phase can
no longer be ignored. The charge conjugation relation of \eqn{dennereqn} 
is seen to
be simply a choice of complex phase. We shall investigate how this 
phase choice relates
to the one used in the so-called
spinor techniques (see, for instance, \cite{ks}). That is, we adopt two
four-vectors ${k_{0,1}}^\mu$ with
\bq
{k_0}^2 = k_0\cdot k_1 = 0\;\;\;,\;\;\;{k_1}^2 = -1
\eq
which we use to define two basis spinors $u_\la$ ($\la=\pm$) :
\bq
u_\la\ubar_\la = {1\over2}\left(\wph1+\la\g^5\right)\fs{k}_0\;\;\;,\;\;\;
u_+ \equiv \fs{k}_1u_-\;\;.
\label{basis}
\eq
For any massless momentum $q$ we can then construct left- and right-handed
spinors as follows :
\bq
u_\la(q) \equiv (2q\cdot k_0)^{-1/2}\,\fs{q}u_{-\la}\;\;\;,\;\;\;
u_\la(q)\ubar_\la(q) = {1\over2}\left(\wph1+\la\g^5\right)\fs{q}\;\;.
\label{masslessfermion}
\eq
Many useful properties of these conventions can be found in \cite{ks} ; 
here, the most important one is reversal :
\bq
\ubar_{\la_1}(q_1)\,\Gamma\,u_{\la_2}(q_2) = \la_1\la_2\;
\ubar_{-\la_2}(q_2)\,\Gamma^R\,u_{-\la_1}(q_1)\;\;,
\label{reversalidentity}
\eq
where $\Gamma$ is any string of Dirac matrices, and 
the superscript ${}^R$ denotes reversal of the order of all Dirac matrices
in $\Gamma$.

We can now define spinors for massive fermions as well. Let $p^\mu$ be the
momentum, and $s^\mu$ the spin vector of such a fermion. Its mass is 
$|m|$, with the convention that $m$ is positive for particles and negative
for antiparticles. Momentum and spin obey
\bq
p^2 = m^2\;\;,\;\;p\cdot s = 0\;\;,\;\;s^2 = -1\;\;\;\Rightarrow\;\;\;
(p\pm ms)^2 = 0\;\;.
\eq
The corresponding spinor can be defined by
\bq
u(\la;p,m,s) = \left(4k_0\cdot(p-\la ms)\right)^{-1/2}\;
\left(\vph\fs{p}+m\right)\left(\vph1+\g^5\fs{s}\right)\,u_\la\;\;;
\eq
some simple Dirac algebra shows that this is equivalent to
\bq
u(\la;p,m,s) \equiv {1\over\sqrt{2}}\left(1+{1\over m}\fs{p}\right)\,
u_{-\la}(p-\la ms)\;\;.
\label{definitionformassivefermion}
\eq
It is easy to check that
\bq
u(\la;p,m,s)\ubar(\la;p,m,s) = \Pi(m,p,s)\;\;,
\eq
as required. There are thus two different phase choices for each 
massive fermion$\;;$ for massless fermions of definite helicity there is
only one nonsingular convention\footnote{For a left-handed fermion,
$p^\mu+ms^\mu$ approaches zero as $m\to0$.}, but at any rate for massless
fermions the spinor defintion (\ref{masslessfermion}) is adequate.

We can now investigate the effect of reversing the fermion flow in a fermionic
current. Let $\Gamma$ be a basis element of the Clifford algebra, that is,
\bq
\Gamma\;\;\;=\;\;\;
1\;\;,\;\;\g^\mu\;\;,\;\;\si^{\mu\nu}\;\;,\;\;\g^5\g^\mu\;\;,\;\;
\g^5\;\;.
\eq
We write for the current and its flow-reversed form :
\bqa
&&J(\Gamma) = \ubar(\la_1;p_1,m_1,s_1)\,\Gamma\,u(\la_2;p_2,m_2,s_2)\;\;,\nl
&&J(\Gamma)^F = \la_1\la_2\;\;
\ubar(-\la_2;p_2,-m_2,s_2)\,\Gamma\,u(-\la_1;p_1,-m_1,s_1)\;\;.
\eqa
Note that $J^F$ is written using $\Gamma$ and {\em not\/} $\Gamma^R$.
Using the reversal identity of \eqn{reversalidentity} and the spinor
definition of \eqn{definitionformassivefermion}, it is simple to
obtain relations between $J$ and $J^F$. As an example, we consider $\Gamma=1$ 
with $\la_1=\la_2=\la$ :
\bqa
J(1) &=& \ubar(\la;p_1,m_1,s_1)\,u(\la;p_2,m_2,s_2)\nl
&=& N\;\ubar_{-\la}(p_1-\la m_1s_1)
\left(1+{1\over m_1}\fs{p}_1\right)
\left(1+{1\over m_2}\fs{p}_2\right)u_{-\la}(p_2-\la m_2s_2)\nl
&=& N\;\ubar_{-\la}(p_1-\la m_1s_1)
\left({1\over m_1}\fs{p}_1+{1\over m_2}\fs{p}_2\right)
u_{-\la}(p_2-\la m_2s_2)\nl
&=& N\;\ubar_{\la}(p_2-\la m_2s_2)
\left({1\over m_1}\fs{p}_1+{1\over m_2}\fs{p}_2\right)
u_{\la}(p_1-\la m_1s_1)\nl
&=& -\;N\;\ubar_{\la}(p_2-\la m_2s_2)
\left(1-{1\over m_2}\fs{p}_2\right)
\left(1-{1\over m_1}\fs{p}_1\right)u_{\la}(p_1-\la m_1s_1)\nl
&=& - \ubar(-\la;p_2,-m_2,s_2)\,u(-\la;p_1,-m_1,s_1) = - J(1)^F\;\;,
\eqa
where
\bq
N = \left(4k_0\cdot(p_1-\la_1m_1s_1)\right)^{-1/2}
\left(4k_0\cdot(p_2-\la_2m_2s_2)\right)^{-1/2}\;\;.
\eq
Note that changing the sign of both $\la$ and $m$ preserves the value of $N$.
The above reasoning is based on the fact that between $\ubar_\la$ and $u_\la$
only an {\em odd\/} number of Dirac matrices survives, while between
$\ubar_\la$ and $u_{-\la}$ only an {\em even\/} number of Dirac matrices
gives a nonzero result. 
Moreover it must be realized that, for the basis spinors,
\bq
\left(\wph u_\la\ubar_\la\right)^R = u_{-\la}\ubar_{-\la}\;\;\;,\;\;\;
\left(\wph u_\la\ubar_{-\la}\right)^R = - u_\la\ubar_{-\la}\;\;.
\eq
Using all this, it is simple to derive the following relations :
\bq
\begin{tabular}{lclclcl}
$J(1)$ &=& $ -J(1)^F$ &$\hphantom{AAAAA}$&  $J(\g^\mu)$&=&$ J(\g^\mu)^F$\\
$J(\g^5\g^\mu)$&=&$ -J(\g^5\g^\mu)^F$ &$\hphantom{AAAAA}$
& $J(\si^{\mu\nu})$&=&$ J(\si^{\mu\nu})^F$\\
$J(\g^5)$&=&$ -J(\g^5)^F$
\end{tabular}\label{propertiis}
\eq
It must be noted that the identities (\ref{propertiis}) can also be proven
directly by explicit traces, for instance for $\la_1=\la_2$ we have
\[
J(\Gamma)\;\;\propto\;\;\tr{\left(1+\la_1\g^5\right)\fs{k}_0
\Pi(m_1,p_1,s_1)\,\Gamma\,\Pi(m_2,p_2,s_2)}
\]
but proving the identities then involves using the Schouten
identity\footnote{In four dimensions, $
g^{\la\mu}\,\eps^{\nu\ro\al\be} +
g^{\la\nu}\,\eps^{\ro\al\be\mu} +
g^{\la\ro}\,\eps^{\al\be\mu\nu} +
g^{\la\al}\,\eps^{\be\mu\nu\ro} +
g^{\la\be}\,\eps^{\mu\nu\ro\al} = 0$.} in many places.
Now, the flow reversal implies the interchange of two fermions
and ought to introduce a minus sign in the flow-reversed current.
The scalar, axial and pseudoscalar currents behave
`correctly', while flow reversal must be accompanied by a sign
change for vector and tensor vertices : this accords with
\cite{denner}. We conclude that the correct way of implementing the effect
of flow reversal for external fermions is
\bq
u(\la;p,m,s)\;\;\to\;\;\la\;\ubar(-\la;p,-m,s)
\eq
and vice versa. As to the internal Dirac structure, the rule is simply
that every Dirac matrix undergoes a sign change, and the string is written
in reversed order. In that case, \eqn{propertiis} naturally arises, and
also any fermionic propagator with numerator $\fs{q}+m$ is transformed
into $-\fs{q}+m$, in agreement with the prescription that the momentum
should be counted in the direction of the fermion flow.
Note that applying flow reversal {\em twice\/} brings $u(\la;p,m,s)$
not back to itself but rather to $-u(\la;p,m,s)$. Since this holds
for both external fermions in a fermionic current, this does not lead to
any problems ; but it indicates that the charge conjugation assignment
implied by the definition of \eqn{definitionformassivefermion} is
non-trivial.\\

In closing, we want to stress that the spinor-antispinor transformation
for which the charge conjugation matrix is commonly employed can also be
obtained without it. Let us consider a spinor $\xi$ for an on-shell particle,
given as simply four complex numbers in a scheme where we know the vectors
$k_0$ and $k_1$ that were employed to construct it using the above spinor
techniques. Therefore there exist $\la$, $m$, $p$ and $s$ such that
$\xi=u(\la;p,m,s)$. It is easily seen that $\bar{\xi}\xi=2m$, 
$\bar{\xi}\g^\mu\xi=2p^\mu$ and $\bar{\xi}\g^5\g^mu\xi=-2ms^\mu$, so that
$m$, $p$ and $s$ are readily determined. Furthermore, $\la$ can be unearthed
from the fact that $u_\la\xi$ is always {\em real\/} while $u_{-\la}\xi$
is always\footnote{Except in the singuklar case where either $p+ms$ or
$p-ms$ happens to be parallel to $k_0$.} complex. We can therefore
construct $u(-\la;p,-m,s)$ without any reference to the $C$ matrix.
Nevertheless, in practical calculations (see for instance \cite{oord})
$C$ is probably the more computation-efficient tool.


\end{document}